\def\bsax{{\it Beppo}SAX\ }

\def\hetetwo{{\it HETE-2}\ }
\def\hetetwonosp{{\it HETE-2}}

\def\ojet{\Omega_{\rm jet}}
\def\oview{\Omega_{\rm view}}
\def\eop{E^{\rm obs}_{\rm peak}}
\def\eiso{E_{\rm iso}}

\def\ep{E_{\rm peak}}
\def\se{S_E}
\def\fn{F^{\rm P}_N}
\def\einf{E^{\rm inf}_{\gamma}}
\def\egt{E^{\rm true}_{\gamma}}

\def\thetav{\theta_{\rm v}}
\def\thetan{\theta_{0}}
\def\thetaj{\theta_{\rm jet}}
\def\egamma{E_{\gamma}}
\def\epei{\ep \propto \eiso^{1/2}}
\def\epeioff{\ep \propto \eiso^{1/3}}

      [1999/12/01 v1.4c Il Nuovo Cimento]
\documentclass{cimento}

\usepackage{graphicx}
\title{Jet Models of X-Ray Flashes}
\author{D. Q.~Lamb\from{uofc},
T. Q.~Donaghy\from{uofc} \&
C.~Graziani\from{uofc}}
\instlist{\inst{uofc} Dept. of Astronomy \& Astrophysics, 
University of Chicago\\ 5640 S. Ellis Ave., Chicago, IL, 60615}
\PACSes{\PACSit{98.70.Rz}{gamma-ray sources; gamma-ray bursts}}
\newcommand{\beq}{\begin{equation}}
\newcommand{\eeq}{\end{equation}}
\begin{document}

\maketitle

\begin{abstract}
One third of all HETE-2--localized bursts are X-Ray Flashes (XRFs), a
class of events first identified by Heise in which the fluence in the
2-30 keV energy band exceeds that in the 30-400 keV energy band.  We
summarize recent HETE-2 and other results on the properties of XRFs. 
These results show that the properties of XRFs, X-ray-rich gamma-ray
bursts (GRBs), and GRBs form a continuum, and thus provide evidence
that all three kinds of bursts are closely related phenomena.  As the
most extreme burst population, XRFs provide severe constraints on burst
models and unique insights into the structure of GRB jets, the GRB
rate, and the nature of Type Ib/Ic supernovae.  We briefly mention a
number of the physical models that have been proposed to explain XRFs. 
We then consider two fundamentally different classes of
phenomenological jet models: universal jet models, in which it is
posited that all GRBs jets are identical and that differences in the
observed properties of the bursts are due entirely to differences in
the viewing angle; and variable-opening angle jet models, in which it
is posited that GRB jets have a distribution of jet opening angles and
that differences in the observed properties of the bursts are due to
differences in the emissivity and spectra of jets having different
opening angles.  We consider three shapes for the emissivity as a
function of the viewing angle $\thetav$ from the axis of the jet: 
power-law, top hat (or uniform), and Gaussian (or Fisher).  We then
discuss the effect of relativistic beaming on each of these models.  We
show that observations can distinguish between these various models. 
\end{abstract}

\begin{figure}[htb]
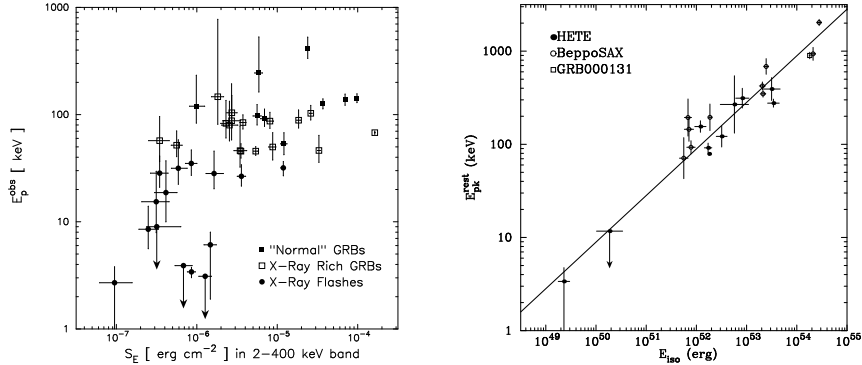

\begin{center}
\rotatebox{270}{\resizebox{4.8cm}{!}{\includegraphics{f1.eps}}}
\qquad
\rotatebox{270}{\resizebox{4.8cm}{!}{\includegraphics{fg2b.ps}}}
\end{center}
\caption{
Left panel: Distribution of HETE-2 bursts in the [$S(2-400
{\rm keV}), E^{\rm obs}_{\rm peak}$]-plane, showing XRFs (filled
circles), X-ray-rich GRBs (open boxes), and GRBs (filled boxes).  This
figure shows that XRFs and X-ray-rich GRBs comprise about 2/3 of the
bursts observed by HETE-2, and that the properties of the three kinds
of bursts form a continuum.  All error bars are 90\% confidence level. 
From \cite{sakamoto2004b}.  Right panel:  Distribution of HETE-2 and
\bsax bursts in the ($\eiso$,$\ep$)-plane, where $\ep$ is the energy of
the peak of the  burst $\nu F_\nu$ spectrum in the source frame.  The
HETE-2 bursts confirm the relation between $\eiso$ and $\ep$ found by
\cite{amati2002} and extend it by a factor $\sim 300$ in $\eiso$.  From
\cite{lamb2004}.
}
\vspace{-0.2truein}
\end{figure}

\vspace{-0.4truein}
\section{Introduction}

Results from HETE-2 show that the properties of XRFs \cite{heise},
X-ray-rich GRBs, and GRBs form a continuum in the [$S_E(2-400~{\rm
kev}),\eop$] plane \cite{sakamoto2004b} (see Figure 1, left panel). 
They also show that the relation between the isotropic-equivalent burst
energy $\eiso$ and the peak energy $\ep$ of the burst spectrum in $\nu
F_\nu$ in the rest frame of the burst found by \cite{amati2002} extends
to XRFs and X-ray-rich GRBs(XRRs) \cite{lamb2005} (see Figure 1, right
panel).  A key feature of the distribution of bursts in these two
planes is that the density of bursts is roughly constant along these
relations, implying equal numbers of bursts per logarithmic interval in
$S_E$, $\eop$, $\eiso$ and $\ep$.  These results, when combined with
earlier results \cite{heise,kippen2002}, strongly suggest that all
three kinds of bursts are closely related phenomena.  

However, the nature of XRFs remains largely unknown.  Key unanswered
questions include the following:  Are $\egamma$ (XRFs) $\ll$ $\egamma$
(GRBs)?  Is the XRF population a direct extension of the GRB and XRR
populations?  Are XRFs a separate emission component of GRBs?  Are XRFs
due to different physics than GRBs?  Finally, does the burst population
extend down to the UV and optical energy bands?

As the most extreme burst population, XRFs provide severe constraints
on burst models.  In this paper, we review recent theoretical efforts
to provide a unified jet picture of XRFs, XRRs, and GRBs, motivated by
HETE-2 observations of these three kinds of bursts.  We show that
observations of XRFs can provide unique insights into the structure of
GRB jets, the GRB rate, and the nature of Type Ib/Ic supernovae.

\section{Physical Jet Models of XRFs}

\begin{table}[b]
\centerline{\bf Phenomenological Jet Models of XRFs Considered in This Review}
\begin{tabular}{lcr}
\hline\hline
Jet Profile & Jet Opening Angle & References \\
\hline
Power-law & Universal &
\cite{firmani2004,granot2003,ldg2005,liang2004b,nakar2004,norris2002,psf2003,rossi2004,zhang2002} \\
Power-law + Beaming & Universal & \cite{d2005,gld2005} \\
\hline
Uniform & Universal & ------- \quad \\
Uniform + Beaming & Universal & \cite{d2005,gld2005,yamazaki2002,yamazaki2003} \\
Uniform & Variable &
\cite{bloom2003,frail2001,granot2003,ldg2005,nakar2004,norris2002,rossi2004}\\
Uniform + Beaming & Variable &
\cite{d_rome,d2005,gld2005,yamazaki2004} \\
\hline
Gaussian/Fisher & Universal &
\cite{dz2004,dlg_rome,dlg2005,firmani2004,gld2005,lloyd-ronning2003,zhang2004} \\
Gaussian/Fisher + Beaming & Universal & \cite{d_rome,d2005,gld2005} \\
Gaussian/Fisher & Variable  & \cite{dlg_rome,dlg2005,gld2005,rossi2004} \\
Gaussian/Fisher + Beaming & Variable & \cite{d_rome,d2005,gld2005} \\
\hline\hline
\end{tabular}
\end{table}

\begin{figure}[htb]
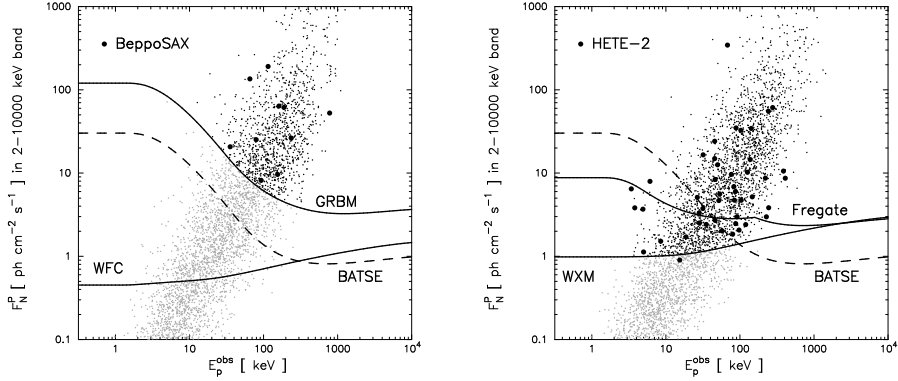

\begin{center}
\rotatebox{270}{\resizebox{5cm}{!}{\includegraphics{Ep_obs_FNP.bsax.m8+data.bw.ps}}}
\qquad
\rotatebox{270}{\resizebox{5cm}{!}{\includegraphics{Ep_obs_FNP.hete.m8+data.bw.ps}}}
\end{center}
\caption{
Distribution of bursts in the [$\eop, \fn(2-10000\ {\rm keV})$]-plane
detected by \bsax (left panel) and by HETE-2 (right panel) compared to
the top-hat VOA jet model.  Detected bursts are shown in black and
non-detected bursts in gray.  For each experiment we show the
sensitivity thresholds for their respective instruments plotted as
solid blue lines.  The BATSE threshold is shown in both panels as a
dashed blue line.  The agreement between the observed and predicted
distributions of bursts is good.  From \cite{ldg2005}.
}
\vspace{-0.15truein}
\end{figure}

A number of theoretical models have been proposed to explain XRFs. 
According to \cite{meszaros2002,zwh2004}, X-ray (20-100 keV) photons
are produced effectively by the hot cocoon surrounding the GRB jet as
it breaks out, and could produce XRF-like events, if viewed well off
the axis of the jet.  However, it is not clear that such a model would
produce roughly equal numbers of XRFs, XRRs, and GRBs, or the
nonthermal spectra exhibited by XRFs.  The ``dirty fireball'' model of 
XRFs posits that baryonic material is entrained in the GRB jet,
resulting in a bulk Lorentz factor $\Gamma$ $\ll$ 300
\cite{dermer1999,huang2002,dermer2003}.  At the opposite extreme, GRB
jets in which the bulk Lorentz factor $\Gamma$ $\gg$ 300 and the
contrast between the bulk Lorentz factors of the colliding relativistic
shells are small can also produce XRF-like events
\cite{mochkovitch2003}.  \cite{yamazaki2002,yamazaki2003,yamazaki2004}
have proposed that XRFs are the result of a highly collimated GRB jet
outside the opening angle of a jet with sharp edges.  In this model,
the low values of $\ep$ and $\eiso$ (and therefore of $\eop$ and $\se$)
seen in XRFs is the result of relativistic beaming.  However, it is not
clear that such a model can produce roughly equal numbers of XRFs,
XRRs, and GRBs, and still satisfy the observed relation between $\eiso$
and $\ep$ \cite{amati2002,lamb2004,lamb2005}.

\begin{figure}[htb]
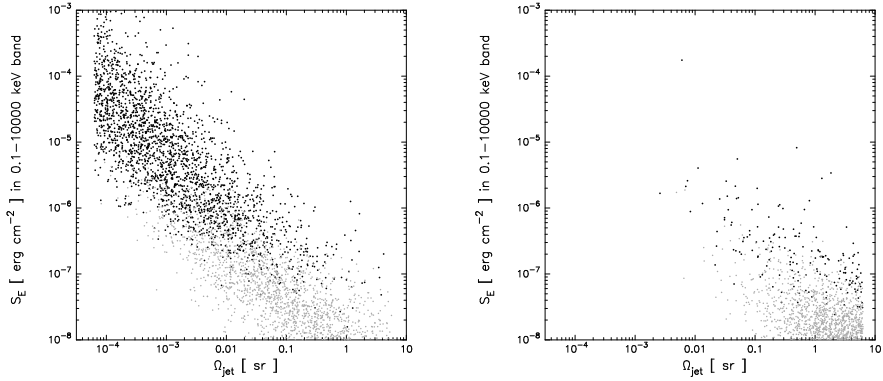

\begin{center}
\rotatebox{270}{\resizebox{5cm}{!}{\includegraphics{omega_SE.m2.bw.ps}}}
\qquad
\rotatebox{270}{\resizebox{5cm}{!}{\includegraphics{omega_SE.univ1m.bw.ps}}}
\end{center}
\caption{
Predicted distribution of bursts in the ($\se$,$\ojet$)-plane in the
top-hat VOA jet model (left panel) and in the PL universal jet model
(right panel).  Bursts detected by the WXM are shown in black and
non-detected bursts in gray.  The left panel exhibits the constant
density of bursts per logarithmic interval in $\se$ and $\ojet$ given
by the top-hat VOA jet model, while the right panel exhibit the
concentration of bursts at $\ojet \equiv \oview \approx 2 \pi$, and the
resulting preponderance of XRFs relative to GRBs, in the PL universal
jet model.  From \cite{ldg2005}.
}
\vspace{-0.15truein}
\end{figure}

\begin{figure}[htb]
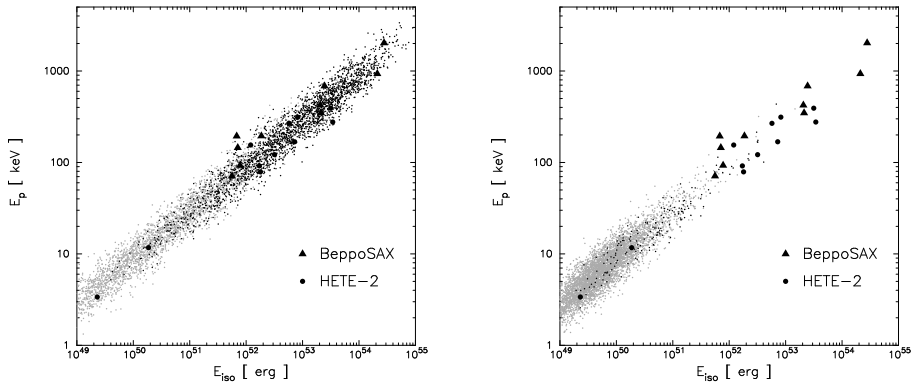

\begin{center}
\rotatebox{270}{\resizebox{5cm}{!}{\includegraphics{Eiso_Ep.m2.bw.ps}}}
\qquad
\rotatebox{270}{\resizebox{5cm}{!}{\includegraphics{Eiso_Ep.univ1m.bw.ps}}}
\end{center}
\caption{
Predicted distribution of bursts in the ($\eiso$,$\ep$)-plane in the
top-hat VOA jet model (left panel) and in the PL universal jet model
(right panel).  Bursts detected by the WXM are shown in black and
non-detected bursts in gray.  The triangles and circles respectively
show the locations of the \bsax and HETE-2 bursts with known
redshifts.  From \cite{ldg2005}.
}
\vspace{-0.15truein}
\end{figure}

\section{Phenomenological Jet Models of XRFs}

In this review, we consider two fundamentally different classes of
phenomenological jet models:  universal jet models, in which it is
posited that all GRBs jets are identical and that differences in the
observed properties of the bursts are due entirely to differences in
the viewing angle; and variable-opening angle (VOA) jet models, in
which it is posited that GRB jets have a distribution of jet opening
angles and that differences in the observed properties of the bursts
are due to differences in the emissivity and spectra of jets having
different jet opening angles.   We consider three shapes for the
emissivity as a function of the viewing angle $\thetav$ from the axis
of the jet:  power-law, top hat (or uniform), and Gaussian (or
Fisher).   We also discuss the effect of relativistic beaming on each
of these models.  Table 1 lists the various models that we consider.

\begin{figure}[htb]
\begin{center}
\raisebox{0.08in}{
\resizebox{6.38cm}{!}{\includegraphics{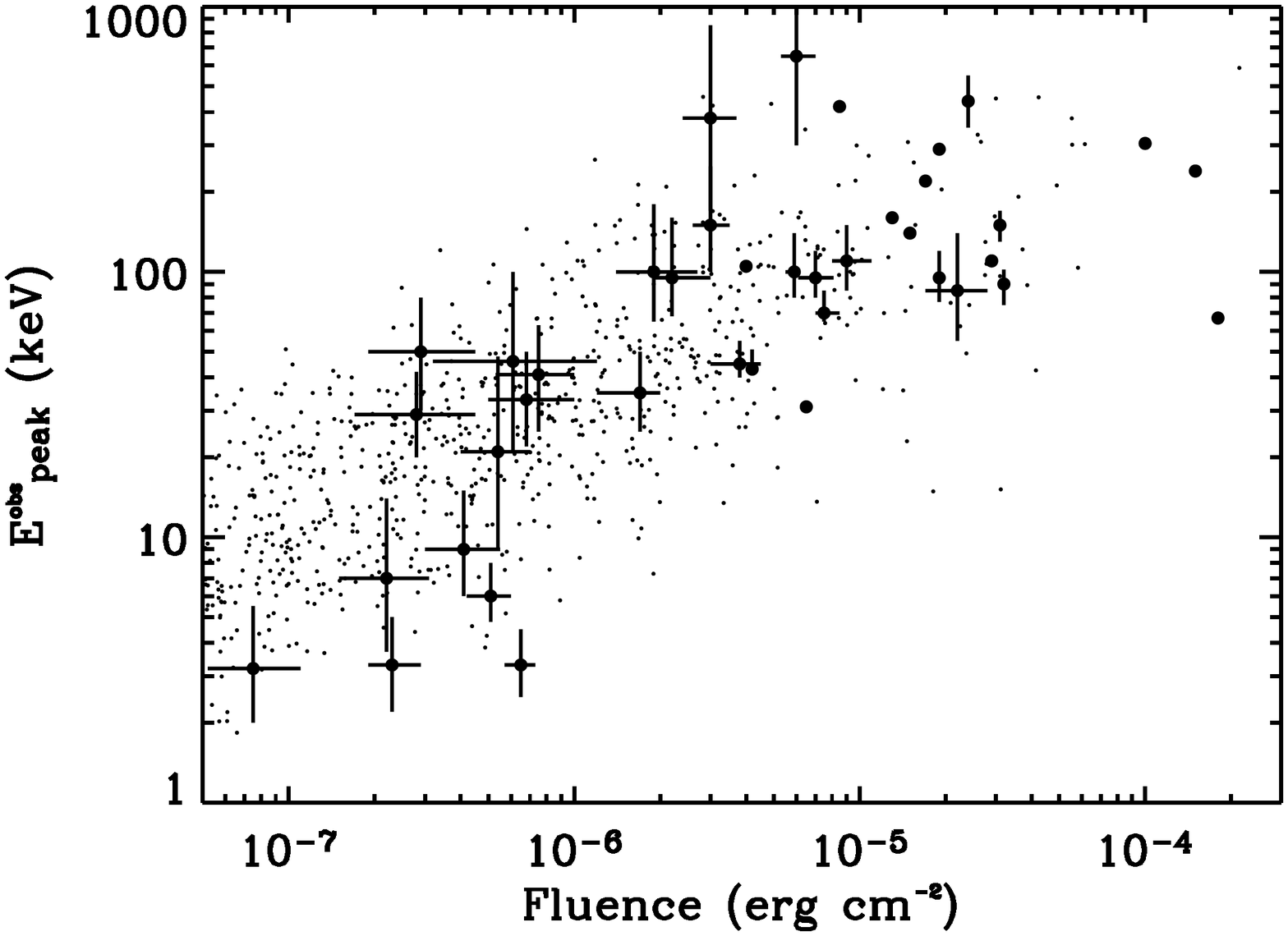}}
}
\quad
\resizebox{6.4cm}{!}{\includegraphics{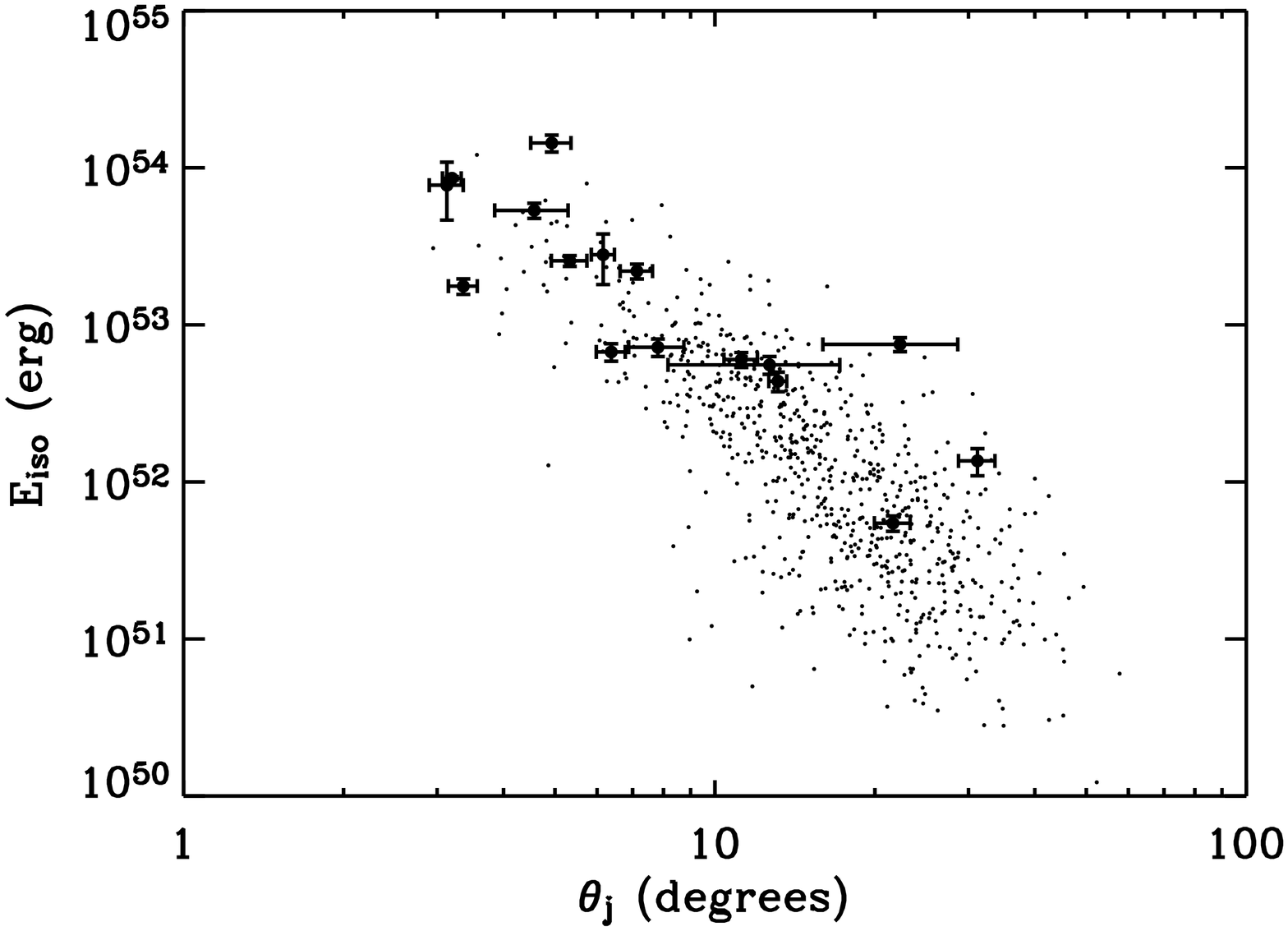}}
\end{center}
\caption{Left panel:  Comparison of the distribution of simulated
bursts in the ($\eop$,$\se$)-plane (small filled circles) for the
Gaussian universal jet model and the distribution of observed HETE-2
and \bsax bursts (large filled circles with error bars; data from
\cite{lamb2004}).  Right panel:  Comparison of the distribution of
simulated bursts in the ($\eiso$,$\thetaj$)-plane (small filled
circles) and the observed distribution (large filled circles with error
bars; data from \cite{bloom2003}, using the afterglow jet break time to
infer $\thetaj$.  Note that the simulated burst distributions are more
strongly concentrated at lower values of $\eop$, $\se$, and $\eiso$,
and larger values of $\thetaj$, than are the observed burst
distributions. From \cite{zhang2004}.}
\vspace{-0.15truein}
\end{figure}

\begin{figure}[htb]
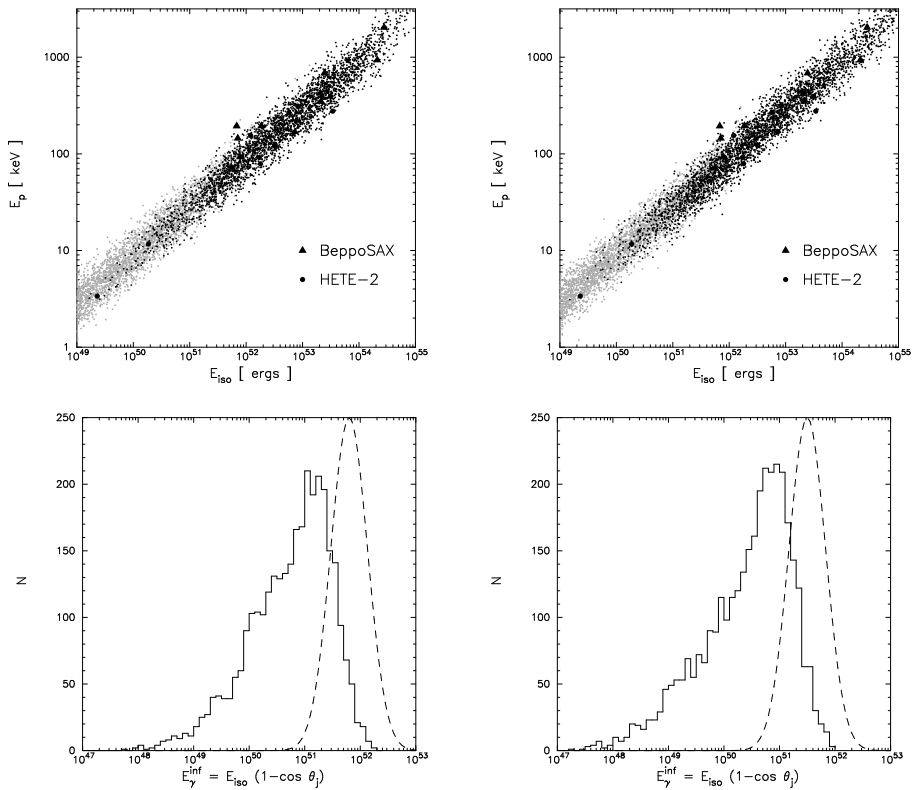

\begin{center}
\rotatebox{270}{\resizebox{5cm}{!}{\includegraphics{donaghy1_f2b.ps}}}
\qquad
\rotatebox{270}{\resizebox{5cm}{!}{\includegraphics{donaghy1_f2c.ps}}}
\end{center}
\begin{center}
\rotatebox{270}{\resizebox{5cm}{!}{\includegraphics{donaghy1_f2e.ps}}}
\qquad
\rotatebox{270}{\resizebox{5cm}{!}{\includegraphics{donaghy1_f2f.ps}}}
\end{center}
\caption{
Top panels:  Predicted distribution of bursts in the 
($\eiso$,$\ep$)-plane in the Fisher universal jet model (left) and in
the Fisher VOA jet model (right).  Bursts detected by the WXM are shown
in black and non-detected bursts in gray.  Bottom panels:  Histograms
of $\einf$ values for detected bursts, as compared to the input
distribution of $\egt$ (dashed curve) for the Fisher universal jet
model (left) and the Fisher VOA jet model (right).  From \cite{dlg2005}
(see also \cite{dlg_rome}).
} 
\vspace{-0.2truein}
\end{figure}

\cite{firmani2004,granot2003,ldg2005,liang2004b,nakar2004,norris2002,psf2003,rossi2004,zhang2002}
consider a universal jet model for GRBs in which the emissivity is a
power-law function of the angle relative to the jet axis.  In
particular, \cite{ldg2005} compare the predictions of a variable
opening angle (VOA) in which the emissivity is roughly uniform across
the head of the jet (see also
\cite{bloom2003,frail2001,granot2003,nakar2004,norris2002,rossi2004})
and a universal jet model in which the emissivity is a power-law
function of the angle relative to the jet axis.  They calculate the
distributions of the observed properties of GRBs, X-ray-rich GRBs, and
XRFs, using Monte Carlo simulations in which they model both the bursts
and the observational selection effects introduced by the \bsax and
HETE-2 instruments (see Figure 2).  They show that, while the uniform
VOA jet model can account for the observed properties of all three
kinds of bursts because of the extra degree of freedom that having a
distribution of jet opening angles gives it, the power-law universal
jet model cannot account for the observed properties of both XRFs and
GRBs (see Figures 3 and 4). 

In response to this conclusion, \cite{dz2004,zhang2004} proposed a
universal Gaussian jet model (see also
\cite{firmani2004,lloyd-ronning2003,zhang2002}). 
They show that such a model can explain many of the observed
properties of XRFs, X-ray-rich GRBs, and GRBs reasonably well (see
Figure 5).  More recently, \cite{dlg_rome,dlg2005,gld2005} considered a
universal jet model in which the emissivity of the jet as a function of
viewing angle is a Fisher distribution (such a distribution is the
natural extension of the Gaussian distribution to a sphere).  They show
that this model has the unique property that it produces equal numbers
of bursts per logarithmic interval in $\eiso$, and therefore in most
burst properties, consistent with the HETE-2 results (see Figure 6, top
row).  They also show that the Fisher universal jet model produces a
broad distribution in the inferred radiated energy $\einf$ (see Figure
6, bottom row).  This is true for any non-uniform jet, since in this
case $\einf = \eiso \cdot (1-\cos \thetaj)$, where $\thetaj = {\rm max}
(\thetan,\thetav)$ \cite{dz2004,kumar2003}, the inferred energy
radiated in gamma rays inferred using the prescription of
\cite{frail2001}, is {\it not} the same as $\egt$, the true energy
radiated in gamma rays.  Thus it is no surprise that when
\cite{dlg_rome,dlg2005} simulate a Fisher VOA jet model, they find
results similar to those for the Fisher universal jet model.

Thus \cite{dlg_rome,dlg2005,gld2005} find that the Fisher universal jet
model and the Fisher VOA jet model make very different predictions for
the distribution of $\einf$ than does the uniform VOA jet model
\cite{ldg2005}.  Further observations of XRFs can determine this
distribution and therefore distinguish between these two models of jet
structure.

\begin{figure}[htb]
\begin{center}
{\resizebox{6.7cm}{!}{\includegraphics{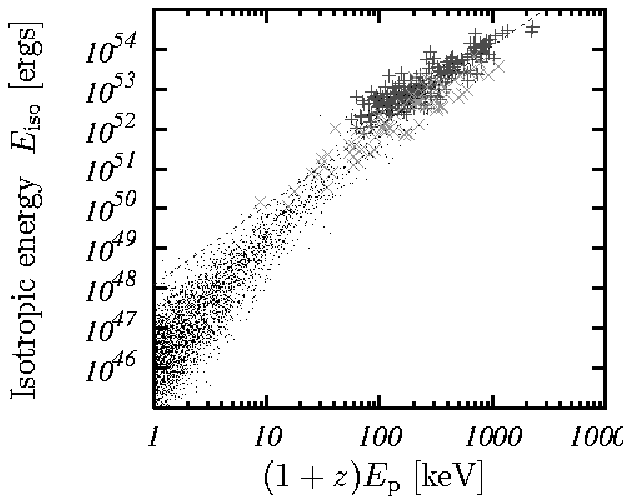}}}
\qquad
\raisebox{0.25in}{
{\resizebox{5cm}{!}{\includegraphics{Eiso_Ep.s16.bw.rot.ps}}}
}
\end{center}
\vspace{-0.2truein}
\caption{Left panel:  Simulation of the distribution of bursts in the
($\ep,\eiso$)-plane for a model in which GRBs are due to uniform VOA
jets for which $\Gamma = 100$ viewed inside $\thetaj$ (crosses) and
XRFs are due to the same jets viewed outside $\thetaj$ (dots).  From
\cite{yamazaki2004}.  Right panel: The same for a model in which both
GRBs and XRFs are due to the variation in the jet opening angle in a
uniform VOA jet model with $\Gamma = 300$.  From \cite{d_rome}.}  
\vspace{-0.2truein}
\end{figure}

\section{Relativistic Beaming}

Relativistic kinematics means that even a ``top-hat''-shaped jet will
be visible when viewed outside of the jet opening angle, $\thetaj$
\cite{ioka2001}.  Relativistic kinematics also means that  $\eiso$ and
$\eop$ will be small in such cases.  \cite{yamazaki2002,yamazaki2003}
used these facts to construct a model in which XRFs are simply
classical GRBs viewed at an angle $\thetav > \thetaj$.  They showed
that such a model could reproduce many of the observed characteristics
of XRFs.; in particular, \cite{yamazaki2004} showed that in such a
model, the distribution of both on- and off-axis observed bursts was
roughly consistent with the $\epei$ relation.

\cite{d_rome,d2005,gld2005} explore further the possibility that the
XRFs observed by \hetetwo and \bsax are primarily off-axis GRBs.  Using
and extending the population synthesis techniques described in
\cite{ldg2005} and \cite{dlg2005}, they present predictions for the
properties of bursts localized by \hetetwonosp.  They show that it is
difficult to account for the observed properties of XRFs by modeling
them solely as ordinary GRBs viewed off-axis.  However, since this
emission must be observable solely on physical grounds, they seek to
understand its relative importance in large burst populations.  They
therefore revisit the uniform VOA jet model put forward in
\cite{ldg2005}, now including the effects of off-axis beaming.  They
note that rough constraints on the bulk $\Gamma$ might be found by
considering the fraction of bursts that are not consistent with the
$\epei$ relation.

\cite{yamazaki2002,yamazaki2003,yamazaki2004} work with a fairly
detailed model of the burst emission; \cite{d_rome} adopts a simpler
model of off-axis beaming in GRB jets.  He makes no assumptions about
the underlying physics generating the burst; instead, he makes the
approximation that the bulk of the emission comes directly from the
edge of the jet closest to the viewing angle line-of-sight (i.e., he
ignores all integrals over the face of the jet and time-of-flight
effects).  His model therefore focuses on the kinematic transformations
of two important burst quantities, $\eiso$ and $\ep$, as a function of
viewing angle.  A more sophisticated treatment of off-axis is
considered in \cite{d2005,gld2005}.

Relativistic kinematics means that frequencies in the rest frame of the
jet will appear Doppler shifted by a factor $\delta^{-1} = \Gamma
(1-\beta \cos\theta)^{-1}$, where $\beta$ is the velocity of the bulk
material and $\theta$ is the angle between the direction of motion and
the source frame observer.  The quantities $\ep$ and $\eiso$ then
transform as $\ep \propto \ep^{\prime} \delta^{-1}$ and $\eiso \propto
\eiso^{\prime} \delta^{-3}$, where $\eiso^{\prime}$ and $\ep^{\prime}$
are the values of these quantities at the edge of the jet.  For a burst
viewed off-axis, these relations imply $\epeioff$.  \cite{yamazaki2004}
do not consider $\eiso$ to be fully bolometric and so derive a slightly
different prescription for the off-axis relation.

Figure 7 compares the relative importance of off-axis beaming for two
different uniform VOA jet models.  The left panel is from 
\cite{yamazaki2004}, who assume $\Gamma=100$ and draw $\thetaj$ values
from a power-law distribution given by $f_{0}\,d\thetaj \propto
\thetaj^{-2}\,d\thetaj$, defined from $0.3$ to $0.03$ rad.  This model
attempts to explain classical GRBs in terms of the variation of jet
opening angles, while XRFs are interpreted as bursts viewed from
outside but close to their jet opening angle.  \cite{d_rome} attempts to
explain both GRBs and XRFs by a distribution of jet opening angles,
following the results presented in \cite{ldg2005}.  He adds off-axis
beaming to this picture.  The right panel shows the results for 
$\Gamma=300$.

\cite{d_rome} finds that the relative importance of off-axis events
increases for models which have very small opening angles.  This is
mainly due to the fact that narrower jets with a constant $\egamma$
will have larger $\eiso$ values, and therefore such bursts viewed
off-axis will also be brighter.  However, Figure 7 (right panel) shows
that the \hetetwo XRFs are not easily explained as classical GRBs
viewed off-axis.  The two XRFs with known redshifts lie along the
$\epei$ relation, and furthermore the larger sample of \hetetwo XRFs
without known redshifts do not fall in the region of the
($\eop$,$S_{\rm E}$)-plane expected for this model: they lie at lower,
rather than higher, $\eop$ values for a given $S_{\rm E}$.  Even given
the model of the off-axis emission in \cite{yamazaki2004}, these
\hetetwo XRFs are difficult to explain.

\cite{d_rome} also considers models that generate
XRFs that obey the $\epei$ relation by extending the range of possible
jet opening-angles to cover five orders of magnitude (see
\cite{ldg2005} for details and discussion).  Hence, XRFs that obey the
$\epei$ relation are bursts that are seen on-axis, but have larger jet
opening angles.  Nonetheless, these models generate a significant
population of off-axis events, although increasing $\Gamma$ reduces the
fraction of off-axis bursts in the observed sample.

\end{document}